\documentclass[12pt]{article}

\usepackage{newlfont}
\usepackage{amsfonts}
\usepackage{amssymb}
\usepackage{amsmath}
\usepackage{latexsym}
\usepackage[all]{xy}

\def\sT{\textsf{T}}
\def\sP{\textsf P}
\def\sV{\textsf V}
\def\R{\mathbb{R}}
\def\I{\mathbf{I}}
\def\Z{\mathbf{Z}}
\def\V{\mathbf{V}}
\def\d{\textsf{d}}
\def\<{\langle} 
\def\>{\rangle} 
\def\({\left(}
\def\){\right)}
\def\p{\mathbf p}
\newtheorem{prop}{Proposition}
\newtheorem{thm}{Theorem}
\newenvironment{pf}{{\noindent{\it Proof. }}}{\ \rule{2mm}{2.5mm}\medskip}
\date{ }

\title{Frame-independent formulation of Newtonian mechanics}

\author{Katarzyna Grabowska \\
{\tt konieczn@fuw.edu.pl}\\
Pawe\l\ Urba\'nski \\
{\tt urbanski@fuw.edu.pl} \\
Physics Department, University of Warsaw\\
Ho\.za 69, 00-681 Warszawa }

\begin{document}

\maketitle

\thanks{Supported by KBN, Grant 2PO3A 041 18}

\begin{abstract}
Based on ideas of W.~M.~Tulczyjew, a frame independent formulation
of analytical mechanics in the Newtonian space-time is presented.
The differential geometry of affine values i.e., the differential
geometry in which affine bundles replace vector bundles and
sections of one dimensional affine bundles replace functions on
manifolds, is used. Lagrangian and hamiltonian generating objects,
together with the Le\-gendre transformation independent on
inertial frame are constructed.

\bigskip\noindent
\textit{MSC 2000: 70G45, 70H03, 70H05}

\medskip\noindent
\textit{Key words: affine spaces, Hamiltonian formalism,
Lagrangian formalism, analytical mechanics}

\end{abstract}

\section{Introduction}
Mathematical formulation of analytical mechanics is usually based
on objects that have vector character. So is the case of the most
of mathematical physics. We use tangent vectors as infinitesimal
configurations, cotangent vectors as momenta, we describe dynamics
using forms (symplectic form) and multivectors (Poisson bracket)
and finally we use an algebra of smooth functions. However, there
are cases where we find difficulties while working with vector
objects. For example, in the analytical mechanics of charged
particles we have a problem of gauge dependence of lagrangians. In
Newtonian mechanics there is a strong dependence on inertial
frame, both in lagrangian and hamiltonian formulation. In the
mechanics of non-autonomous system we are forced to chose a
reference vector field on the space-time that fulfills certain
conditions or we cannot write the dynamics at all. In all those
cases the traditional language of differential geometry seems to
introduce too much mathematical structure. In other words, there
is to much structure with comparison to what is really needed to
define and describe the behavior of the system. As a consequence
we have to put in an additional information to the system such as
gauge or reference frame.

Gauge independence of the Lagrangian formulation of Newtonian
dyna\-mics  can be achieved by increasing the dimension of the
configuration space of the particle.  The four dimensional
space-time of general relativity is replaced by the five
dimensional ma\-nifold (as in the Kaluza theory) (\cite{WMT2},
\cite{DB}). This approach encounters serious conceptual
difficulties and cannot be considered satisfactory. An alternate
approach is proposed in the present note. The four dimensional
space-time is used as the configuration space.  The phase space is
no longer a cotangent bundle and not even a vector bundle.  It is
an affine bundle modelled on the cotangent bundle of the
space-time manifold. The Lagrangian is a section of an affine line
bundle over the tangent bundle of the space-time manifold. The
proper geometric tools are provided by the {\it geometry of affine
values}. We call the geometry of affine values the differential
geometry that is built using sections of one-dimensional affine
bundle over the manifold instead of functions on the manifold. The
affine bundle we use is equipped with the fiber action of the
group $(\R,+)$, so we can add reals to elements of fibres and real
functions to sections, but there is no distinguished "zero
section". Those elements of the geometry of affine values that are
needed in the Newtonian mechanics are described in section 4.1,
the complete presentation of the theory can be found in
\cite{GGU}. In fact the geometry of affine values appeared much
earlier in works of W.M. Tulczyjew and his collaborators (see e.g.
\cite{TUZ}) and has been successfully applied to the description
of the dynamics of charged particles (\cite{TU}).

The aim of this paper is to show how the geometry of affine values works in the simplest
case of analytical mechanics in the Newtonian space-time. The idea of applying affine
geometry to this problem comes from Prof W\l odzimierz Tulczyjew and his group (\cite{WMT1},
\cite{WMT2}, \cite{B}). Another proposition of the frame independent formulation of the
Newtonian mechanics by increasing the dimension of the space-time can be found in
\cite{P}.

\section{Newtonian space-time}

    The {\it Newtonian space-time} is a system $(N,\tau, g)$ where $N$ is a
four-dimensio\-nal affine space with the model vector space $V$, $\tau$ is a
non-zero element of $V^\ast$ and $g\colon E_0\rightarrow E_0^\ast$ represents an
Euclidean metric on $E_0=\ker\tau$. The elements of the space $N$ represent events.
The time elapsed between two events is measured by $\tau$:
    $$\Delta t(x,x')=\<\tau,x-x'\>.$$
    The distance between two simultaneous events is measured by $g$:
    $$d(x,x')=\sqrt{\<g(x-x'),x-x'\>}.$$
    The space-time $N$ is fibrated over the time $T=N\slash E_0 $
which is one-dimensional affine space modelled on $\R$. By $\eta$ we will
denote the canonical projection
        $$\eta\colon N\longrightarrow T, $$
    by $\imath$ the canonical embedding
    $$\imath: E_0 \longrightarrow V ,$$
    and by $\imath^\ast$ the dual projection
    $$\imath^\ast: V^\ast \longrightarrow E_0^\ast.$$
    By means of $\imath$ and $\imath^\ast$ we can define a contravariant tensor $g'$ on
$V^\ast$:
    $$g'=\imath\circ g^{-1}\circ\imath^\ast.$$
    The kernel of $g'$ is a one-dimensional subspace  of $V^\ast$ spanned by $\tau$.

Let $E_1$ be an affine subspace of $V$ defined by the equation $\<\tau,v\>=1$. The
model vector space for this subspace is $E_0$. An element of $E_1$ can represent
velocity of a particle. The affine structure of $N$ allows us to associate to an
element $u$ of $E_1$ the family of inertial observers that move in the space-time
with the constant velocity $u$. This way we can interpret an element of $E_1$ as an
inertial reference frame. For a fixed inertial frame  $u$, we define the space  $Q$
of world lines of all inertial observers. It is the quotient affine space $N/\{u\}$.
The space-time $N$ becomes the product of affine spaces
        $$ N = Q\times T  .
                                                                $$ 
    The model vector space for $Q$ is the quotient vector space $V/\{u\}$ that can be
identified with $E_0$. The corresponding canonical projection is
    $$\imath_u \colon  V\rightarrow E_0\colon  v\longmapsto
\imath_u(v)=v-\<\tau,v\>u $$
    and the splitting $V=E_0\times \R$ is given by
    $$V\ni v\longmapsto (\imath_u(v),\<\tau,v\>)\in E_0\times\R.$$
    The dual splitting is given by
    $$V^\ast \ni p\longmapsto (\imath^\ast(p), \<p,u\>)\in E_0^\ast\times\R.$$

    The tangent bundle $\sT N$ we identify with the product $N\times V$ and the
subbundle  $\sV N$  of vectors  vertical with respect to the
projection on time, with $N\times E_0$. Consequently,  the  bundle
$\sV^1N$ of infinitesimal configurations (positions and
velocities) of particles moving in the space-time $N$ is
identified with $N\times E_1$. When the inertial frame $u$ is
chosen, $E_1$ is identified with $E_0$ and $\sV^1 N$ is identified
with $\sV N$.

The vector dual $\sV^\ast N$ for $\sV N$ is a quotient bundle of $N\times V^\ast$ by
the one-dimensional subbundle $N\times\<\tau\>$. We can identify it  with
$N\times  E_0^\ast$. Using the inertial frame we can make it a subbundle of $\sT^\ast N$.

\section{Analytical mechanics in the fixed inertial frame}

In the following section we will present the analytical mechanics
of one particle in the Newtonian space-time in the fixed inertial
frame $u$. First we concentrate on the inhomogeneous formulation,
suitable for trajectories parameterized by the time, then we pass
to the homogeneous one. The homogeneous formulation accepts all
parameterizations.

\subsection{Inhomogeneous dynamics described in the fixed \\ inertial frame}
\label{inhom}

    Let $u\in E_1$ represent an inertial frame.  For a fixed time $t\in T$, the
phase space for a particle with mass $m$ with respect to the inertial frame $u$ is
$\sT^\ast N_t\simeq N_t\times E_0^\ast$, where $N_t=\eta^{-1}(t)$. The collection
of phase spaces form a phase bundle $\sV^\ast N\simeq N\times E_0^\ast$. Phase
space trajectories of the system are solutions of the well-known equations of
motion:
    \begin{equation}\label{eqmotion}\dot p =-\d_s\varphi(x)\qquad\dot
x=g^{-1}(\frac{p}{m})+u, \end{equation}
    where $(x,p,\dot x,\dot p)\in \sV^1\sV^\ast N \subset\sT\sV^\ast N\simeq
N\times E_0^{\ast}\times V\times E_0^\ast$ and $\varphi\colon N\rightarrow\R$
is a potential. Subscript ${}_s$ in $\d_{s}$ means that we differentiate only
in spatial directions i.e. the directions vertical with respect to the
projection on time, therefore $\d_{s}\varphi(x)\in E_0^\ast$. The equations
define a vector field on $\sV^\ast N$ with values in $\sV^1\sV^\ast N$, i.e.
a section of the bundle $\sV^1\sV^\ast N\rightarrow\sV^\ast N$. The image of
the vector field (\ref{eqmotion}) we will call {\it the inhomogeneous
dynamics} and denote by $D_{i,u}$.
    It can be generated directly by the \textbf{lagrangian}
\begin{equation}\label{inhomlag}
    \ell_{i,u}\colon \sV^1 N \rightarrow \R \colon (x,w)\mapsto
\frac{m}2\<g(w-u),w-u\>-\varphi(x).
 \end{equation}

    The procedure of generating the inhomogeneous dynamics from the lagrangian
(\ref{inhomlag}) is as follows.  The image of the vertical
derivative $\d_s \ell_{i,u}$ is a submanifold of
    \[\sV^\ast \sV^1 N \simeq N\times E_1\times E_0^\ast \times E_0^\ast  \]
which is canonically isomorphic (as an affine space) to
    \[\sV^1\sV^\ast N \simeq N\times  E_0^\ast \times E_1 \times E_0^\ast . \]
    This isomorphism we obtain by a reduction of the canonical isomorphism
    $$\alpha_N \colon \sT\sT^\ast N\rightarrow \sT^\ast \sT N $$
(for the definition of $\alpha_{N}$ see \cite{WMT3}), which for
affine spaces assumes the form
    \[\alpha_N(x,a, v, b) = (x,v, b, a).\label{alf}\]
After the reduction, we get
    \[\alpha_N^1 \colon N\times  E_0^\ast \times E_1 \times E_0^\ast\rightarrow
    N\times E_1\times E_0^\ast \times E_0^\ast. \]
Using $\alpha_N^1$ we can obtain $D_{i,u}$ from $\d_s \ell_{i,u}$
by taking an inverse-image:
    \[D_{i,u}=(\alpha_N^1)^{-1}(\d_s \ell_{i,u}(\sV^1N)).\]

    The dynamics $D_{i,u}$ cannot be generated directly from a hamiltonian
by means of the canonical Poisson structure  on $\sV^\ast N$,
which is the reduced cano\-nical Poisson (symplectic) structure of
$\sT^\ast N$. In the coordinates adapted to the structure of the
bundle $(t,x^i, p_i)$ the Poisson bi-vector is given by
$$\Lambda=\partial p_i\wedge \partial x^i.$$
    Symplectic leaves for this Poisson structure are cotangent bundles $\sT^\ast
N_t$, where $N_t =\eta^{-1}(t)$. It follows that every hamiltonian vector
field is  vertical with respect to the projection on time. However, using
reference frame $u$, we can generate first the vertical part of the dynamics
(\ref{eqmotion}), i.e the equations

    \begin{equation}\label{veqmotion}\dot p =-\d_s\varphi(x)\qquad\dot
x=g^{-1}(\frac{p}{m}), \end{equation}
    and add the reference vector field $u$.
{\bf The hamiltonian function} for the problem reads
$$h_{i,u}(x,p)=\frac{1}{2m}\<p,g^{-1}(p)\>+\varphi(x),$$
where $(x,p)\in\sV^\ast N$.

 The system (\ref{veqmotion}) can be
generated also from lagrangian function defined on $\sV N$ by the formula:
    \begin{equation}
\ell_{i,u}(x,w)=\frac{m}2\<g(w),w\>-\varphi(x).
    \end{equation}
    We identify the fiber over $t$ of $\sV N$ with $\sT N_t$ and use the
standard procedure to generate a submanifold $D_{u,t}$ in $\sT\sT^\ast N_t$.
The collection of these submanifolds give us the system (\ref{veqmotion}).

    The dynamics (\ref{eqmotion}) and the generating procedures depend
strongly on the choice of the reference frame. In particular, the relation
velocity-momenta is frame-dependent which means that we have to redefine the
phase manifold for the particle to obtain frame-independent dynamics.
    Also the hamiltonian formulation will be possible if we replace the
canonical Poisson tensor by a more adapted object.

\subsection{Homogeneous dynamics described in the \\ fixed inertial frame}

    In the homogeneous formulation of the dynamics infinitesimal
configurations are pairs $(x,v)\in N\times V^+ \subset N\times V\simeq \sT N$
where $V^+$ is an open set of vectors  such that $\langle \tau,v \rangle
>0$.

The homogeneous lagrangian is an extension by homogeneity of the $\ell_{i,u}$
from (\ref{inhom}) and is given by the formula:
\begin{equation}
\ell_{h, u}(x,v)=\frac{m}{2\<\tau,v\>}\<g(\imath_u(v)), \imath_u(v)\>-
\<\tau,v\>\varphi(x).
\end{equation}

    This choice guaranties that the action calculated for a piece of the world
line, which is one dimensional oriented submanifold of the
space-time, does not depend on its parametrization. However, we
still have to use the fixed inertial frame $u$.

The image of the differential of $\ell_{h,u}$ is a lagrangian submanifold of
$\sT^\ast\sT N\simeq N\times V\times V^\ast\times V^\ast$. An element
$(x,v,\alpha_x,\alpha_v)$ is in the image of $\d \ell_{h,u}$ if it satisfies
the following equations
\begin{equation}\label{homd}
\left\{
\begin{array}{l}
v\in V^+, \\
\alpha_x=-\<\tau,v\>\d\varphi(x), \\
\alpha_v=\frac{m}{\<\tau,v\>}\imath_u^\ast\circ g\circ \imath_u(v)-
 \frac{m}{2\<\tau,v\>^2}\<g(\imath_u(v)),\imath_u(v)\>\tau
 -\varphi(x)\tau.
\end{array}\right.
\end{equation}
The image of $\d \ell_{h,u}(N\times V^+)$ by the mapping $\alpha_N^{-1}$ is a
lagrangian submanifold of $\sT\sT^\ast N$. This submanifold we will call {\it
the homogeneous dynamics} and denote by $D_{h,u}$. An element $(x,p,\dot
x,\dot p)$ of $\sT\sT^\ast N\simeq N\times V^\ast\times V\times V^\ast$ is in
$D_{h,u}$ if
\begin{equation}\label{homd2}
\left\{
\begin{array}{l}
\dot x=v, \\
\dot p=-\<\tau,v\>\d\varphi(x), \\
p=\frac{m}{\<\tau,v\>}i_u^\ast\circ g\circ \imath_u(v)-
  \frac{m}{2\<\tau,v\>^2}\<g(\imath_u(v)),\imath_u(v)\>\tau
 -\varphi(x)\tau
\end{array}\right.
\end{equation}
for some $v\in V^+$, i.e. $\<\tau,v\> > 0$. We observe that $D_{h,u}$ does
not project on the whole $\sT^\ast N$, but $(x,p)$ must satisfy the following
equation:
\begin{equation}
\frac{1}{2m}\<p,g'(p)\>+\<p,u\>+\varphi(x)=0. \label{kmu}
\end{equation}
The equation (\ref{kmu}) is the analog of the mass-shell equation
$p^2=m^2$ in the relativistic mechanics. Since there is the
difference in signature of $g'$ between the Newtonian and the
relativistic case, we obtain here paraboloid of constant mass
instead of relativistic hyperboloid. The mass-shell will be
denoted by $K_{m,u}$.

  It is possible  to  generate the dynamics  $D_{h,u}$ directly  by a
generalized hamiltonian system. The hamiltonian generating object (see
\cite{TU2}) is the family

\begin{equation} \label{fam1}\xymatrix@C=2cm{
N\times V^\ast\times V^+ \ar[r]^-{-H_{h,u}} \ar[d]_\zeta& \R\\
N\times V^\ast & },\end{equation}
    where
\begin{equation}
H_{h,u}(x,p,v)=\<p,v\>-\ell_{h,u}(x,v)\in \R.
\end{equation}

    This family can be simplified. The fibration $\zeta$ can be represented
as a composition $\zeta''\circ \zeta'$, where
    \[\zeta'\colon N\times V^\ast\times V^+ \rightarrow N\times V^\ast\times\R_+
    \colon (x,p,v)\mapsto (x,p,<\tau,v>), \]
and
     \[\zeta''\colon N\times V^\ast\times \R_+ \rightarrow N\times V^\ast
    \colon (x,p,r)\mapsto (x,p). \]
    Equating to zero the derivative of $H_{h,u}$ along the fibres of $\zeta'$
    we obtain the relation

\begin{equation}
    v = \frac{<\tau,v>}{m} g^{-1}\circ \imath^\ast (p) + <\tau,v>u.
\end{equation}

    It follows that the family (\ref{fam1}) is equivalent (generates the same
object) to the reduced family
    \begin{equation} \label{fam2}\xymatrix@C=2cm{
N\times V^\ast\times \R_+ \ar[r]^-{-\widetilde{H}_{h,u}} \ar[d]_-{\zeta''}& \R\\
N\times V^\ast & },\end{equation}
    where
    \begin{equation}
\widetilde{H}_{h,u}(x,p,r)=r(\frac{1}{2m}\<p,g'(p)\>+\<p,u\>+\varphi(x)).
\end{equation}
    No further simplification is possible.

    The critical set $S(\widetilde{H}_{h,u}, \zeta''))$ is the submanifold
    \[\left\{(x,p,r)\in N\times V^\ast\times \R_+ ;\quad \frac{1}{2m}\<p,g'(p)\>
    +\<p,u\>+\varphi(x)=0\right\}\]
    and its image $\zeta''(S(\widetilde{H}_{h,u}, \zeta''))$ is the mass shell
$K_{m,u}$.

    The function $H_{h,u}$ is zero on $S(\widetilde{H}_{h,u}, \zeta'')$ and projects to
the zero function on $K_{m,u}$. However, a Dirac system with the
zero function on the constraints $K_{m,u}$ does not generate
$D_{h,u}$. The lagrangian submanifold  $\bar{D}_{h,u}\subset
\sT\sT^\ast N$ generated by this system is exactly the
characteristic distribution of $K_{m,u}$, i.e.
$$\bar{D}_{h,u}=(\sT K_{m,u})^\S$$ and does not respect the condition
$\langle\tau,v\rangle>0$. We have only $D_{h,u}\subsetneq \bar{D}_{h,u}$.

\section{The dynamics independent on inertial\\ frame}

\subsection{The geometry of affine values}

First, let us give some definitions. A vector space $V$ with
distinguished non-zero element $v$ we will call a {\it special
vector space}. A canonical example of a special vector space is
$(\R,1)$. It will be denoted by $\I$. If $A$ is an affine space
then Aff$(A,\R)$ -- the vector space of all affine functions with
real values on $A$ -- is a special vector space with distinguished
element $1_A$ being a constant function on $A$ equal to 1. The
space Aff$(A,\R)$ will be denoted by $A^\dagger$ and called a {
\it vector dual} for $A$. Having a special vector space $(V,v)$ we
can define its {\it affine dual} by choosing a subspace in
$V^\ast$ of those linear functions that take the value $1$ on $v$:
$$V^{\ddagger}=\{\varphi\in V^\ast:\quad \varphi(v)=1\}.$$
We have that
\begin{thm}[\cite{GGU}]
For $(V,v)$ and $A$ such that \emph{dim}~$V < \infty$ and
\emph{dim}~$A < \infty$
$$\begin{array}{c}
\left((V^\ddagger)^\dagger, 1_{V^\ddagger}\right)=V, \\
\left(A^\dagger\right)^\ddagger=A.
\end{array}$$
\end{thm}
An affine space modelled on a special vector space will be called a {\it
special affine space}. Similar definitions we can introduce for bundles: a
{\it special vector bundle} is a vector bundle with distinguished
non-vanishing section and a {\it special affine bundle} is an affine
bundle modelled on a special vector bundle.

The geometry of affine values is, roughly speaking, the differential
geometry built on the set of sections of one-dimensional special affine
bundle $\Z$ over $M$ modelled on $M\times \I$, instead of just functions
on $M$. The bundle $\Z$ will be called a {\it bundle of affine values}.
Since $\Z$ is modelled on $M\times\I$ we can add reals in each fiber of
$\Z$, i.e $\Z$ is  an $\R$-principal bundle. The vertical vector field on
$\Z$ which is the fundamental vector field for the action of $\R$ will be
denoted by $X_\Z$. Let us now consider an example of a bundle of affine
values: If $(V,v)$ is a special vector space, then we have the quotient
vector space $V_0=V\slash \<v\>$. The vector spaces $V$ and $V_0$ together
with the canonical projection form an example of a bundle $\V$ of affine
values. The appropriate action of $\R$ in the fibers is given by
$$ V\times\R\ni(w,r)\longmapsto (w-rv)\in V$$
and the fundamental vector field $X_\V$ is a constant vector field equal to
$v$ on $V$.

The affine analog of the cotangent bundle $\sT^\ast M$ in the
geometry of affine values is called a {\it phase bundle} and
denoted by $\sP \Z$.  We define an equivalence relation in the set
of pairs of $(m,\sigma)$, where $m\in M$ and $\sigma$ is a section
of $\Z$. We say that $(m,\sigma)$, $(m',\sigma')$ are {\it
equivalent} if $m=m'$ and $\d(\sigma-\sigma')(m)=0$, where we have
identified the difference of sections of $\Z$ with a function on
$M$. The equivalence class of $(m,\sigma)$ is denoted by
$\d\sigma(m)$. The set of equivalence classes is denoted by
$\sP\Z$ and called the {\it phase bundle} for $\Z$. It is, of
course, the bundle over $M$ with the projection
$\d\sigma(m)\mapsto m$. It is obvious that $\sP\Z\rightarrow M$ is
an affine bundle modelled on the cotangent bundle $\sT^\ast
M\rightarrow M$.

As an example we construct a phase bundle for the bundle of affine values
built out of a special vector space. In the set of all sections of the
bundle $\V$ there is a distinguished set of affine sections, since $V$ and
$V_0$ as vector spaces are also affine spaces. We observe that there are
affine representatives in every equivalence class $\d\sigma(m)$ that
differ by a constant function. There is also one linear representative,
i.e. such an affine section that takes value $0$ at the point $0\in V_0$.
The set of elements of a phase bundle can be therefore identified with a
set of pairs: point in $m$ and a linear injection from $V_0$ to $V$.
Moreover, we observe that such linear injections are in one-to-one
correspondence with linear functions on $V$ such that they take value $1$
on $v$ (or the canonical vector field $X_\V$ evaluated on the function
gives $1$). The image of a linear section is a level-$0$ set of the
corresponding function. The functions that correspond to linear sections
form the affine dual $V^\ddagger$, therefore we have
\begin{equation}\label{dag}
  \sP\V\simeq V_0\times V^\ddagger.
\end{equation}

\subsection{Frame independent lagrangian}

Now, we will collect all the homogeneous lagrangians for all inertial frames
and construct for them a universal  object which does not depend on an
inertial frame. It is convenient to  treat a lagrangian as a section of the
trivial bundle $N\times V\times \R\rightarrow N\times V$ rather than as a
function.

For two reference frames $u$ and $u'$, we have the  difference
    $$\ell_{h,u}(x,v)-\ell_{h,u'}(x,v)=m\<g(u'-u),\imath_{\frac{u'+u}{2}}(v)\>.$$
Let us denote  $\imath_{\frac{u'+u}{2}}^\ast g(u'-u)$ by $\sigma(u',u)$. With
this notation
$$\ell_{h,u}(x,v)-\ell_{h,u'}(x,v)=m\<\sigma(u',u),v\>. \label{lla}$$
For $\sigma$ we have the following equalities
\begin{equation}
\sigma(u',u)=-\sigma(u,u'), \label{sa}
\end{equation}
\begin{equation}
\sigma(u'',u')+\sigma(u',u)=\sigma(u'',u). \label{sb}
\end{equation}

In the $E_1\times N\times V\times\R$, we introduce the following relation:
\begin{equation}
(u,x,v,r)\sim (u',x',v',r')\quad\Longleftrightarrow\quad\left\{
\begin{array}{l}
x=x',\\
v=v',\\
r=r'+m\<\sigma(u',u),v\>.
\end{array}\right.\label{row}
\end{equation}
From (\ref{sa}) we obtain that $\sim$ is symmetric and reflexive,
from (\ref{sb}) that it is transitive, therefore it is an
equivalence relation. Since the relation does not affect $N$ at
all, it is obvious that in the set of equivalence classes we have
a cartesian product structure $N\times W$. In $W$ we distinguish
two elements: $w_0=[u,0,0]$ and $w_1=[u,0,-1]$,
$$w_0=\{(u,0,0):\,\, u\in E_1\},\qquad w_1=\{(u,0,-1):\,\, u\in E_1\},$$
and two natural operations:
$$+: W\times W\rightarrow W\qquad \circ:\R\times W\rightarrow W$$
\begin{equation}
\begin{array}{l}
[u,v,r]+[u',v',r']= \\
\quad=[\frac{u+u'}{2},\, v+v',\,
r+r'+m(\langle\sigma(u,\frac{u+u'}{2}),v\rangle+
\langle\sigma(u',\frac{u+u'}{2}),v'\rangle)], \\
\alpha\circ[u,v,r]=[u,\alpha v, \alpha r].
\end{array}
\end{equation}
The above operations are well defined that can be checked by direct
calculation. Some more calculation one needs to show that

\begin{prop}
$(W,+,\circ)$ is a vector space with $w_0$ as the zero-vector. Moreover
$(W,w_1)$ is a special vector space such that $W\slash<w_1>\simeq V$
\end{prop}

\noindent The canonical projection $W\rightarrow V$ will be denoted by
$\zeta$.
    It follows from \ref{lla} that quadruples $(u,x,v,\ell_{h,u}(x,v))$ and
    $(u',x,v,\ell_{h,u'}(x,v))$ are equivalent. Consequently, frame
 dependent lagrangian defines a section $\ell_h$ over $N\times V^+$ of  the
one-dimensional special affine bundle (a bundle of affine values) $N\times
W\rightarrow N\times V$ which does not depend on the inertial frame.
 The section $\ell_h$ will be called an {\it
affine lagrangian} for the homogeneous mechanics independent on
the choice of inertial frame. In the following we show that the
bundle $N\times W\rightarrow N\times V$ carries a structure, which
can be used for generating the frame-independent dynamics. We
begin with the construction of the phase space.

\subsection{Phase space}

In the  frame dependent formulation of the dynamics,  the phase space for the
massive particle is $\sT^\ast N\simeq N\times V^\ast$. For each frame $u$ we
have the Legendre mapping
\begin{equation}\label{leg}
\begin{array}{rcl}
  \cal L_u &\colon &\sT N \supset N\times V^+\rightarrow \sT^\ast N \\
   &\colon &(x,v) \mapsto
\frac{m}{\<\tau,v\>}\imath_u^\ast\circ g\circ \imath_u(v)-
 \frac{m}{2\<\tau,v\>^2}\<g(\imath_u(v)),\imath_u(v)\>\tau -\varphi(x)\tau,
\end{array}
\end{equation}
    i.e. the vertical derivative of $\ell_{h,u}$ with respect to the projection $\sT^\ast
    N \rightarrow N$.

Since $\ell_{h,u}(x,v)-\ell_{h,u'}(x,v)=m\<\sigma(u',u),v\>$, we have also
\begin{equation} \label{leg2}
  \cal L_u(v) -\cal L_{u'}(v) = m\sigma(u',u).
\end{equation}

\begin{prop}
A mapping $\Phi_{u',u}\colon \mathsf T^\ast N \rightarrow \mathsf T^\ast N$
defined by
    \[\Phi_{u',u}(x,p)= (x, p + m\sigma(u',u))  \]
    has the following properties
\begin{enumerate}
  \item $\Phi_{u',u}(K_{m,u'})= K_{m,u}$,
  \item it is a symplectomorphism of the canonical symplectic structure  on
  $\mathsf T^\ast N$,
  \item $\mathsf T \Phi_{u',u}(D_{h,u'}) = D_{h,u}$.
\end{enumerate}
\end{prop}
\begin{pf} \label{leg3}
    The image of $\cal L_u$ is $K_{m,u}$, so the first property is an
    immediate consequence of  (\ref{leg2}) and the definition of $\Phi_{u',u}$.
    The mapping $\Phi_{u',u}$ is a translation by a constant vector. It
    follows that it is a symplectomorphism. Consequently,
    \[\sT \Phi_{u',u}((\sT K_{m,u'})^\S) = (\sT K_{m,u})^\S\] and
    \[\sT \Phi_{u',u}(\bar{D}_{h,u'} ) = \bar{D}_{h,u}.\]
    Since  $\Phi_{u',u}$ respects the time orientation, we have also
       \[\sT \Phi_{u',u}({D}_{h,u'} ) = {D}_{h,u}.\]
\end{pf}

 The above observation suggests the
following equivalence relation in $E_1\times N\times V^\ast$:

\begin{equation}
(u,x,p)\sim (u',x',p')\quad\Longleftrightarrow\quad\left\{
\begin{array}{l}
x=x',\\
p=p'+m\sigma(u',u).
\end{array}\right.
\end{equation}
Again, we have the obvious structure of the cartesian product in the set of
equivalence classes: $N\times P$. The set $N\times P$ will be called an {\it
affine phase space}.  The set $P$ is an affine space modelled on $V^\ast$:
$$[u,p]+\pi=[u, p+\pi]\text{ for }\pi\in V^\ast.$$
An element of $P$ will be denoted by $\p$.

It follows from Proposition~\ref{leg3} that $N\times P$ is a symplectic
manifold and the isomorphism of tangent and cotangent bundles assumes the
form
\begin{equation}\label{beta}
\begin{array}{rl}
 \beta &\colon \sT (N\times P)\simeq N\times P\times V\times
 V^\ast\longrightarrow \sT^\ast (N\times P)\simeq N\times P\times V^\ast\times
 V \\
    &\colon(x,\p,v,a) \longmapsto (x,\p, a, -v)
\end{array}
\end{equation}
 Moreover, the equivalence classes of the elements of mass-shells form the
universal mass shell $K_m$ and the elements of frame dependent dynamics form
the universal dynamics $D_h$  which is contained in $(\sT K_{m})^\S$.

    A straightforward calculation shows that  the function
        \[E_1\times N\times V^\ast\ni (u,x,p)\mapsto \frac{1}{2m}\<p,g'(p)\>+\<p,u\>\]
is constant on equivalence classes and projects to a function on
$N\times P$. We denote this function by $\Psi_m$. It follows that
the generating object (\ref{fam2}) of the dynamics $D_{h,u}$
defines a generating object
  \begin{equation} \label{fam3}\xymatrix@C=2cm{
N\times P\times \R_+ \ar[r]^-{-\widetilde{H}_{h}} \ar[d]_-{\zeta''}& \R\\
N\times V^\ast & },\end{equation}
    of the dynamics $D_h$, where
        \begin{equation}
\widetilde{H}_{h}(x,p,r)=r(\Psi_m+\varphi(x)).
\end{equation}
\subsection{Lagrangian as a generating object}
    In the previous section we have constructed the frame independent
dynamics $D_h$ and a hamiltonian generating object. Now, we show
that the frame independent affine lagrangian $\ell_h$ is also a
generating object of $D_h$. $\ell_h$ is a section of a bundle of
affine values $N\times W \rightarrow N\times V$ and its
differential is a section of $\sP (N\times W) \rightarrow N\times
V$ over $N\times V^+$. For a given frame $u$, we identify $N\times
W$ with $N\times V\times \R$ and a section of $\zeta$ with a
function on $N\times V$. Consequently, an affine  covector $a\in
\sP (N\times W)$ is represented by a covector $a_u\in
\sT^\ast(N\times V)= N\times V \times V^\ast \times V^\ast$. It
follows from (\ref{row}) that  $a_u= (x,v, a,b)$  and
$a_{u'}=(x,v,a,b +m\sigma(u,u'))$ represent the same element of
$\sP (N\times W)$. In the process of generation of frame dependent
dynamics we use the canonical isomorphism $\alpha_N  \colon
\sT\sT^\ast N\rightarrow \sT^\ast \sT N$ (\ref{alf}). We observe
that
    $$\alpha_N (x,a +m\sigma(u,u') ,v,b) = (x,v,b,a +m\sigma(u,u')), $$
hence $\alpha_N$ defines an isomorphism
    $$\alpha \colon \sT(N\times P)\rightarrow \sP(N\times W)$$
    and the image of $(D_h)$ is the image of $\d l_h $.

    Now, we can summarize our constructions. We have canonical symplectic
    structure on $N\times P$ with the corresponding mapping
$$\beta:\sT(N\times P)\longrightarrow\sT^\ast(N\times P),$$
    which forms the basis for the hamiltonian formulation of the dynamics.
    Together with $\alpha$ it gives rise to the following diagram (Tulczyjew
    triple):

$$\xymatrix@C-30pt{
(\sT^\ast(N\times P), \omega_{N\times P})\ar[dr] &
 &
(\sT (N\times P), d_{\sT}\omega_P) \ar[ll]_{\beta}\ar[rr]^{\alpha} \ar[dr]
\ar[dl] &
 &
(\sP(N\times W),\omega_{N\times W}) \ar[dl]\\
& N\times P & & N\times V & }
$$

\subsection{The Legendre transformation}

    The Legendre transformation is the passage from lagrangian to hamiltonian
generating object. In previous sections it was done with the
knowledge of the Legendre transformation for the frame dependent
dynamics. In that case we make use of the canonical
symplectomorphism $\gamma_M\colon \sT^\ast\sT^\ast M \rightarrow
\sT^\ast \sT M$ generated by $\langle \,,\,\rangle \colon \sT
M\times_M \sT^\ast M\rightarrow \mathbb{R}$, where $M$ is a
manifold and $\langle \,,\,\rangle$ is the canonical pairing
between vectors and covectors. It follows that the image
$\gamma(L)$ of a lagrangian submanifold $L$ generated by a
lagrangian $\ell$ is generated by a Morse family
    $$  \ell-\langle \,,\,\rangle \colon \sT M\times_M \sT^\ast M\rightarrow
    \mathbb{R},$$
    where $\sT M\times_M \sT^\ast M $ is considered a fibration over
    $\sT^\ast M$
    (see \cite{TU2} for details).

    Now, we show that analogous procedure can be applied in the case of
the affine framework.
    First, we observe that every element  $w\in W$ defines, in natural way,
an affine function on $P$:
\begin{equation}
f_w(\p)=\<p,v\>-r,\quad\text{where}\quad w=[u,v,r],\,\,\p=[u,p].
\end{equation}
Indeed, when we take another  representative of $w$ and $\p$, e.g.
$(u',v,r')$ and $(u',p')$ respectively, then we obtain
$$\<p',v\>-r'=\<p-m\sigma(u',u),v\>-r+m\<\sigma(u',u),v\>=\<p,v\>-r.$$ The element $w_1$
defines the constant function equal to $1$ on $P$:
$$f_{w_1}(p)=\<p,0\>-(-1)=1.$$
This implies the following:
\begin{prop}
There is a natural isomorphism between $P^\dag$ and $(W,w_1)$ given by
$$f_{[u,v,r]}([u,p])=\<p,v\>-r.$$
It means that also $W^{\ddag}\simeq P$.
\end{prop}
    With this isomorphism we have (see (\ref{dag})) $\sP(N\times W) \simeq N\times
    V\times V^\ast \times P$ and $\alpha \colon \sT(N\times P)\rightarrow
    \sP(N\times W)$ assumes the form
\begin{equation}\label{alph}
 \alpha \colon (x,\p,v, a)\longmapsto (x,v,a,\p).
\end{equation}

\noindent $(W,w_1)$, being a special vector space, has a structure of a
one-dimensional affine bundle modelled on $V\times\mathbf I$. The action
of the group $(\R, +)$ in the fiber over $V$ comes from the natural action
in the fiber of $E_1\times V\times \R\rightarrow E_1\times V$. The
fundamental vector field $X_W$ for this action is a constant vector field
with value $w_1$ at every point.

Now, we need a pairing  between $P$ and $V$, which reduces to $\langle
\,,\,\rangle$ (as a section of the trivial bundle $V\times V^\ast \times
\mathbb{R}$) in the vector case. The pairing is a section of $P\times W$ over
$P\times V$ defined by
\begin{equation}
P\times V\ni (\p,v)\longmapsto \<\p,v\>=[u,v,\<p,v\>]\in W,\quad
\text{where}\,\, \p=[u,p].
\end{equation}
The above definition is correct, i.e. does not depend on the choice of
representatives:
\begin{equation}\label{pair}
  [u,v,\<p,v\>]=[u',v,\<p,v\>-\<m\sigma(u',u),v\>]=[u',v,\<p',v\>].
\end{equation}

It remains to show that the pairing (\ref{pair}) generates an isomorphism
between $\sT (N\times P)$ and $\sP (V\times W)$.

\begin{prop}
There is a natural symplectomorphism between $\mathsf{P}((N\times
W)\times(N\times P))$ and $\mathsf{P}(N\times W)\ominus
\mathsf{T}^\ast(N\times P)$.
\end{prop}
    \begin{pf}
    It is enough to check that any section of $(N\times W)\times(N\times
P)$ over $(N\times V)\times(N\times P)$ is equivalent to a section $\sigma$
of the form
    $$\sigma(x,v,y,\p) =  \sigma_0(v) + f_1(x) - f_2(\p) -f_3(y), $$
where $\sigma_0$ is a linear section of $W\rightarrow V$ and functions $f_i$
are affine.
    \end{pf}

    Similar arguments show  that $\sP(N \times W)\simeq N\times V\times V^\ast
    \times P$.

    The canonical diagonal inclusion $N\subset N\times N$ implies the
projection
    $$V^\ast\times V^\ast \rightarrow V^\ast\colon (a,b)\mapsto a+ b$$
and consequently a relation between $\sP (N\times W \times P)$ and $\sP
((N\times W) \times (N\times P))$. With this relation a section of $\sP
(N\times W \times P)$ over $N\times V \times P$ defines a submanifold of
    $$\sP ((N\times W) \times (N\times P))= \sP(N\times W)\ominus \sT^\ast(N\times
    P)$$
i.e., a symplectic relation
    $$  \sT^\ast(N\times P)\longrightarrow\sP(N\times W).$$
    In particular, the differential of the pairing $\<\,,\,\>$ generates a
    relation

\begin{equation}
    \gamma \colon  \sT^\ast(N\times P)\simeq N\times P\times V^\ast \times
    V\longrightarrow\sP(N\times W) \simeq N\times V\times V^\ast\times P
\end{equation}
   It easy task to verify that this relation has the following representation
    $$N\times P\times V^\ast \times V \ni (x,\p,a, v) \longmapsto (x, -v, a, \p)
     \in N\times V\times V^\ast\times P.$$

    We see from (\ref{alph}) and (\ref{beta}) that $\gamma =\alpha\circ
    \beta^{-1}$, and  consequently $\gamma\circ\alpha(D_h) = \beta (D_h)$.
    Following the general rule for composing of generating objects,  we
    conclude that $\beta (D_h)$ is generated by the Morse family

\begin{equation}
  \label{fam4}\xymatrix@C=2cm{
N\times P\times V^+ \ar[r]^-{-H_{h}} \ar[d]_-{\zeta''}& \R\\
N\times P&},
\end{equation}
    where
\begin{equation}
  H_h(x,v,\p) =  \< \p,v \> -\ell_h(x,v).
\end{equation}
    As in the frame-dependent case, this family can be reduced to the family
    (\ref{fam3}).

\subsection{Comments}
    Frame independent inhomogeneous formulation of the dynamics can be
    obtained by the reduction of the homogeneous lagrangian with respect to
    the embedding $N\times V^1 \hookrightarrow N\times V$. The Legendre
    transformation leads to a hamiltonian which is no longer a function, but
    a section of certain bundle over $N\times P$. It requires also extension of the notion
    of a Poisson bracket to sections of an affine values bundle (see
    \cite{GGU}, \cite{PU}).

\end{document}